\documentclass[twocolumn,twocolappendix]{aastex631}
\usepackage{amsmath, amssymb}
\usepackage{graphicx,color,float,tabularx}
\usepackage[caption=false]{subfig}
\usepackage{placeins}
\usepackage{rviewport}
\definecolor{colorLink}{rgb}{0,0,180} 
\usepackage{hyperref}
\hypersetup{
   colorlinks = true,
   citecolor  = colorLink,
   urlcolor   = colorLink,
   linkcolor  = colorLink,
}

\newcommand{\paper}{paper}

\begin{document}

\title{Neutrino anisotropy as a probe of extreme astrophysical accelerators}

\author[0000-0003-4615-5529]{Marco Stein Muzio}
\email{muzio@wisc.edu}
\affiliation{Department of Physics and Wisconsin IceCube Particle Astrophysics Center, University of Wisconsin, Madison, WI, 53706, USA}
\affiliation{Department of Physics, Pennsylvania State University, University Park, PA 16802, USA}
\affiliation{Department of Astronomy and Astrophysics, Pennsylvania State University, University Park, PA 16802, USA}
\affiliation{Institute of Gravitation and the Cosmos, Center for Multi-Messenger Astrophysics, Pennsylvania State University, University Park,
PA 16802, USA}

\author[0000-0001-9011-0737]{No\'{e}mie Globus}
\affiliation{Instituto de Astronomía, Universidad Nacional Autónoma de México Campus Ensenada, A.P. 106, Ensenada, BC 22860, México}
\affiliation{Kavli Institute for Particle Astrophysics and Cosmology, Stanford University, Stanford, CA 94305, USA}
\affiliation{Astrophysical Big Bang Laboratory, RIKEN, Wako, Saitama, Japan}

\date{\today}

\begin{abstract}
We predict that neutrino sources following the matter distribution of the universe result in an anisotropy in the neutrino sky imprinted by the local large-scale structure. We calculate the level of this anisotropy and explore how it depends on the cosmological evolution of neutrino sources. We show how the level of anisotropy can be amplified when a cutoff in the neutrino spectrum is considered, introducing an effective neutrino horizon. This effect might allow for future neutrino detectors to measure a neutrino anisotropy associated with the local large-scale structure. Measurement of the level of this anisotropy along with features of the neutrino spectrum will allow observers to constrain the cosmological evolution of neutrino sources, which at ultrahigh energies (UHEs) are also expected to be the sources of UHE cosmic rays.  
\end{abstract}

\section{Introduction}
\par
The discovery of a diffuse neutrino flux has opened a new window on the universe. The diffuse neutrino spectrum has been observed from ${\sim}10$~TeV to ${\sim}10$~PeV by IceCube~\citep{IceCube:2023wmh,Naab:2023xcz} and extended by a single $220^{+570}_{-110}$~PeV observation by KM3NeT~\citep{KM3NeT:2025npi}. Below roughly $30$~PeV, cosmic rays produced by a variety of Galactic and extragalactic accelerators can contribute to this spectrum. Above roughly $30$~PeV only cosmic rays produced by the most extreme accelerators, ultrahigh energy cosmic rays (UHECRs, $E_{CR} \gtrsim 1$~EeV), have sufficient energy to contribute to the diffuse neutrino spectrum.\footnote{The highest energy neutrinos produced by CRs are those produced photohadronically with energy of roughly $E = E_\mathrm{CR}/A/20$. Neutrinos produced by CRs hadronically have much lower energies in general. Therefore, a $30$~PeV neutrino requires a primary CR with an energy-per-nucleon of at least $10^{17.7}$~eV. Such large energies-per-nucleon are only carried by extragalactic CRs at UHEs.} This implies that the neutrino spectrum could continue beyond EeV energies, since the CR spectrum has been observed up to ${\sim}100$~EeV. 

\par
Up to now, no significant neutrino anisotropy or correlation with source populations has been found~\citep{Ackermann:2022rqc,IceCube:2020svz,IceCube:2021waz,IceCube:2022ccm,IceCube:2019cia,IceCube:2022der,IceCube:2022zbd,IceCube:2019yml,IceCube:2021slf,IceCube:2021imv}, though one point source, NGC 1068~\citep{IceCube:2022der}, and a correlation with the Galactic plane have been detected at lower energies~\citep{IceCube:2023ame}. 

\par
At high redshifts, the matter distribution of the universe, or the large-scale structure (LSS), is highly isotropic. However, locally, at smaller redshift scales, the LSS is highly anisotropic~\citep{Hoffman:2018ksb}. This raises the possibility that an anisotropy may exist in the neutrino sky if one assumes that the distribution of astrophysical neutrino sources follows the LSS. Further, an anisotropy is guaranteed to exist above some energy threshold as long as the neutrino spectrum cuts off at high energies and neutrino sources exist within the volume of the local LSS, since neutrinos at near-cutoff energies must originate from low redshifts. Throughout this paper we primarily focus on astrophysical neutrinos (those produced inside astrophysical sources) rather than cosmogenic neutrinos (those produced in extragalactic propagation), since neutrinos produced inside large-scale structures are dominantly astrophysical. We note that such a cutoff is expected empirically, regardless of the site of neutrino production, because at the highest energies only UHECRs can produce neutrinos and the UHECR spectrum itself has been observed to cutoff around $10^{19.7}$~eV~\citep{PierreAuger:2020kuy}. 

\par
In particular, high energy neutrinos provide a unique probe of UHECRs, whose sources remain an open question in astroparticle physics~\citep{Anchordoqui:2018qom,AlvesBatista:2019tlv,Coleman:2022abf,Globus23}. UHECRs themselves probe only the most local sources (within ${\sim}100$~Mpc at $100$~EeV for nuclei $A\leq 56$, e.g.~\citet{Ding:2021emg}) due to horizons imprinted by the Greisen-Zatsepin-Kuzmin (GZK) effect~\citep{Greisen:1966jv,Zatsepin:1966jv} and time delays from extragalactic magnetic fields~\citep{Lipari:2008zf,Murase:2008sa,Globus:2018svy}. Moreover, UHECRs which do reach Earth undergo large deflections in the Galactic magnetic field~\citep{Pshirkov:2013wka,Golup:2009cv,Farrar2019}, making source identification extremely challenging, even for the most extreme events~\citep{2023ApJ...945...12G}. High energy neutrinos provide a complementary window into UHECR sources since they both point back to their sources and probe cosmological distances. 

\par
In this \paper{}, we show that, in addition to small-scale anisotropies explored by previous studies~\citep{Ahlers:2014ioa,Fang:2016hop,Fiorillo:2022ijt}, a large-scale anisotropy is imprinted by the local LSS and propose that it provides a powerful new tool for investigating the origin of diffuse neutrinos. 

\section{Model}\label{sec:model}
\par
Recent studies (e.g.~\citet{Harari:2013pea,Harari:2015hba,Tinyakov:2014fwa,diMatteo:2017dtg,Globus:2017fym,Ding:2021emg,Allard:2021ioh,Bister:2023icg}) have shown that the UHECR dipole is likely to originate from the large-scale structure (LSS) of the local universe, whose imprint on the sky is then transformed by Galactic magnetic field lensing effects. This result depends on two key insights: first, that the distribution of UHECR accelerators should follow the matter distribution; and second, that the extragalactic background light (EBL), cosmic microwave background (CMB), and intervening magnetic fields introduce a horizon to the UHECR-observable universe. This horizon ensures that the local LSS is the most prominent structure in the UHECR sky, rather than the homogeneous, isotropic universe at higher redshifts.\footnote{We note that the assumption that UHECR accelerators trace the matter distribution does not imply every galaxy, in particular the Milky Way, is host to such an accelerator.}

\par
Neutrinos and their sources also have both of these properties. Neutrino sources should also follow the matter distribution. In particular, astrophysical neutrinos are produced when CRs interact with ambient gas and photons at acceleration sites or in their source environments [see, e.g.,~\citet{Unger:2015laa,Harari:2016vtz,Kachelriess:2017tvs,Fang:2017zjf,Anchordoqui:2018qom,Condorelli:2022vfa}]. This implies that if UHECR accelerators follow the matter distribution of the universe, then so will astrophysical neutrino sources.\footnote{A common tracing of the matter distribution among UHECR and neutrino sources is expected even if the properties of environments host to UHECR accelerators are such that sources are never both UHECR-bright and neutrino-bright.} 

\par
In addition, an effective neutrino horizon can be generated either by the intrinsic distribution of neutrino sources or by observers. If neutrinos are dominantly produced at low redshifts the lack of sources at high redshifts would limit the neutrino-observable universe in the same way as a true horizon. Similarly, an observer integrating a neutrino spectrum with maximum energy $E_\mathrm{max}$ above an energy threshold $E_\mathrm{th}$ can induce a neutrino horizon, since all neutrinos beyond a redshift $z_\mathrm{max} = E_\mathrm{max}/E_\mathrm{th} - 1$ will arrive at Earth with energies below this threshold. For example, if $E_\mathrm{th}=0.9E_\mathrm{max}$, then neutrinos will only probe sources within $z<z_\mathrm{max} \simeq 0.11$ (roughly $460$~Mpc comoving distance). If $E_\mathrm{max}$ is a soft cutoff energy (e.g.\ the energy scale in an exponential cutoff), then this horizon will also be soft. To account for both of these effects, we quantify the redshift of this effective horizon, $z_\mathrm{eff}$, as the redshift within which $90\%$ of the observed neutrino flux originates. In both cases, this effective horizon is generally much larger than that for UHECRs. 

\par 
Therefore, it is plausible that the neutrino sky also has an anisotropy imprinted by the local LSS. If neutrino sources are negatively evolving (decreasing luminosity-density with redshift) observers will measure the neutrino sky to be highly anisotropic, following the local matter distribution as in the case of UHECRs. By contrast, if the neutrino source evolution is positive (increasing luminosity-density with redshift), this anisotropy will be much weaker since the relative flux from the local LSS will be suppressed. We will show that the source evolution sets a minimum level of anisotropy. This anisotropy can then be amplified by observers tuning their energy threshold to probe an arbitrarily small volume of the universe, provided that a sufficiently large detector and the existence of neutrino sources within this local volume.

\par
To quantify the level of anisotropy we make a few simplifying assumptions. First, we assume all sources emit a common neutrino spectrum, $Q_\nu(E)$, and that these sources are the dominant contribution to the observed neutrino flux. Evidence for the former assumption has been found for UHECR source spectra~\citep{Ehlert:2022jmy}, which would then be imprinted onto their neutrino secondaries, but a discussion exploring the case of multiple source populations can be found in Section~\ref{sec:nonStdSrcs}. Second, we assume that the neutrino source density at low redshifts follows the spatial distribution of the local LSS matter density, whereas at high redshifts it is nearly isotropic. Importantly while the relative flux from a given direction and redshift depends on the matter density, the total flux from a given redshift depends on the source luminosity-density (i.e.\ the source evolution), $\mathcal{H}(z)$.

\par
To realistically model the distribution of the local LSS we use the quasi-linear density field from CosmicFlows-2, which is based on an ensemble of $20$ constrained realizations of the local (within $360$~Mpc) universe~\citep{Hoffman:2018ksb}. A view of this median density field is shown at \href{https://skfb.ly/6AFxT}{https://skfb.ly/6AFxT}. The density field beyond the box boundaries is obtained in the linear regime using a series of constrained linear realizations [based on the linear WF/CRs algorithm~\citep{HoffmanRibak91,Zaroubi:1998di} within a $1830$~Mpc depth. The use of the linear realizations is justified as the contributions to the anisotropy from beyond the $360$~Mpc box are dominated by large (linear) scales [see~\citet{Hoffman:2018ksb} for more details]. Beyond $1830$~Mpc we assume the density field fluctuations are small enough to be well-approximated as a perfectly isotropic contribution. Importantly, CosmicFlows-2 only models the local LSS and does not model individual galaxies, which could result in smaller-scale anisotropies. However, modeling at this level of granularity is not necessary since here we are only interested in calculating the diffuse neutrino background associated with the local LSS.   

\par
The total neutrino flux $J_\nu$ above an energy threshold $E_\mathrm{th}$ is given by

\begin{align}\label{eq:totalFlux}
    J_\nu(E\geq E_{th}) = \frac{c}{4\pi} \int_0^{\infty} \frac{dz}{H(z)} \int_{E_{th}}^\infty \mathcal{H}(z) Q_\nu((1+z)E) dE~,
\end{align}

\noindent
adopting the notation of~\citet{Ahlers:2014ioa}. A detailed derivation of this formula can be found in~\citet{Lipari:2008zf}. For clarity, we note that in equation~\eqref{eq:totalFlux} we take $Q_\nu(E) = 1/L_\nu dN_\nu(E)/dE/dt$ to be the spectrum emitted by a source relative to its total luminosity in neutrinos $L_\nu$ (in units of $1/\mathrm{erg}/\mathrm{GeV}$) and $\mathcal{H}(z) = L_\nu(z) dn_s/dz$ to be the total luminosity-density of sources at redshift $z$ (in units of $\mathrm{erg}/\mathrm{s}/\mathrm{Mpc}^{3}$). This ensures that the energy and redshift dependence of the total spectral emission rate density, $\mathcal{L}_\nu(E,z) = \mathcal{H}(z) Q_\nu(E)$, factorizes with $\mathcal{H}$ quantifying the overall normalization of neutrino emission at a given redshift and $Q_\nu$ its energy distribution. 

\par
In order to calculate the total neutrino anisotropy one needs to determine how this total flux is distributed on the sky. Adopting a \textsc{healpix} pixelization~\citep{Gorski:2004by} with $N_\mathrm{side} = 64$ (corresponding to a pixel resolution of roughly $1^\circ$), we obtain the neutrino intensity skymap by calculating the total neutrino flux in each pixel $i$, $J_{\nu,i}$. 

\par
The fraction of flux, $f_j$, originating from a spherical shell, $j$, spanning from redshift $z_j$ to $z_j+\Delta z_j$ is given by

\begin{align}
    f_j = \frac{c}{4\pi J_\nu} \int_{z_j}^{z_j + \Delta z_j} \frac{dz}{H(z)} \int_{E_{th}}^\infty \mathcal{H}(z) Q_\nu[(1+z)E] dE~.
\end{align}

\noindent
Within this spherical shell, the neutrino flux will be distributed according to the matter distribution. If the total mass within shell $j$ is $M_j$ and the total mass within the voxel subtended by pixel $i$ within shell $j$ is $M_{ij}$, then the voxel will produce $M_{ij}/M_j$ of the shell's total flux. Within $1830$~Mpc comoving distance we calculate this ratio using the CosmicFlows-2 database. Beyond $1830$~Mpc, we assume the shell's mass is isotropically distributed so that $M_{ij} = M_j \Omega_{i}/4\pi$, where $\Omega_{i}$ is the solid angle subtended by pixel $i$. 

\par
The flux within pixel $i$ is then given by summing over each shell's contribution

\begin{align}\label{eq:pixelFlux}
    J_{\nu,i} = J_\nu \displaystyle\sum_j f_j \frac{M_{ij}}{M_j}~.
\end{align}

\par
In order to calculate \eqref{eq:pixelFlux}, one needs the neutrino source spectrum $Q_\nu(E)$ and the source evolution $\mathcal{H}(z)$. In practice, since neutrinos are only affected by redshift losses, the parameters of their spectrum at Earth can be directly related to the parameters of their source spectrum. In particular, the source's spectral index is the same as that observed on Earth at energies sufficiently below the observed cutoff energy. The observed cutoff energy can be up to $5$ times lower than that of the source, depending on how positive the source evolution is and how hard its spectrum is. However, for a given observed spectral index and source evolution one can straight-forwardly relate the observed cutoff energy to that of the source. For illustrative purposes, we adopt a single power-law spectrum with an exponential cutoff

\begin{align}
    Q_\nu (E) \propto E^{\gamma} e^{-E/E_\mathrm{max}}~. 
\end{align}

\noindent
Importantly, the absolute scale of $E_\mathrm{max}$ does not matter for the purposes of calculating the neutrino anisotropy, only the ratio $E_\mathrm{th}/E_\mathrm{max}$.

\par
For simplicity, we use a two-parameter model of the source evolution

\begin{align}\label{eq:evo}
    \mathcal{H}_{m,z_0}(z) &= \mathcal{H}_0
    \begin{cases} 
      (1+z)^m & z \leq z_0 \\
      (1+z_0)^m e^{-(z-z_0)} & z > z_0
   \end{cases}~,
\end{align}

\noindent
where the evolution peaks at $z_0$ for $m>0$. This model is an adequate approximation to many source evolutions obtained from astrophysical observations and allows us to explore the range of plausible source evolutions in a straightforward way. 

\par
\begin{figure}[htpb!]
    \centering
    \includegraphics[width=\linewidth]{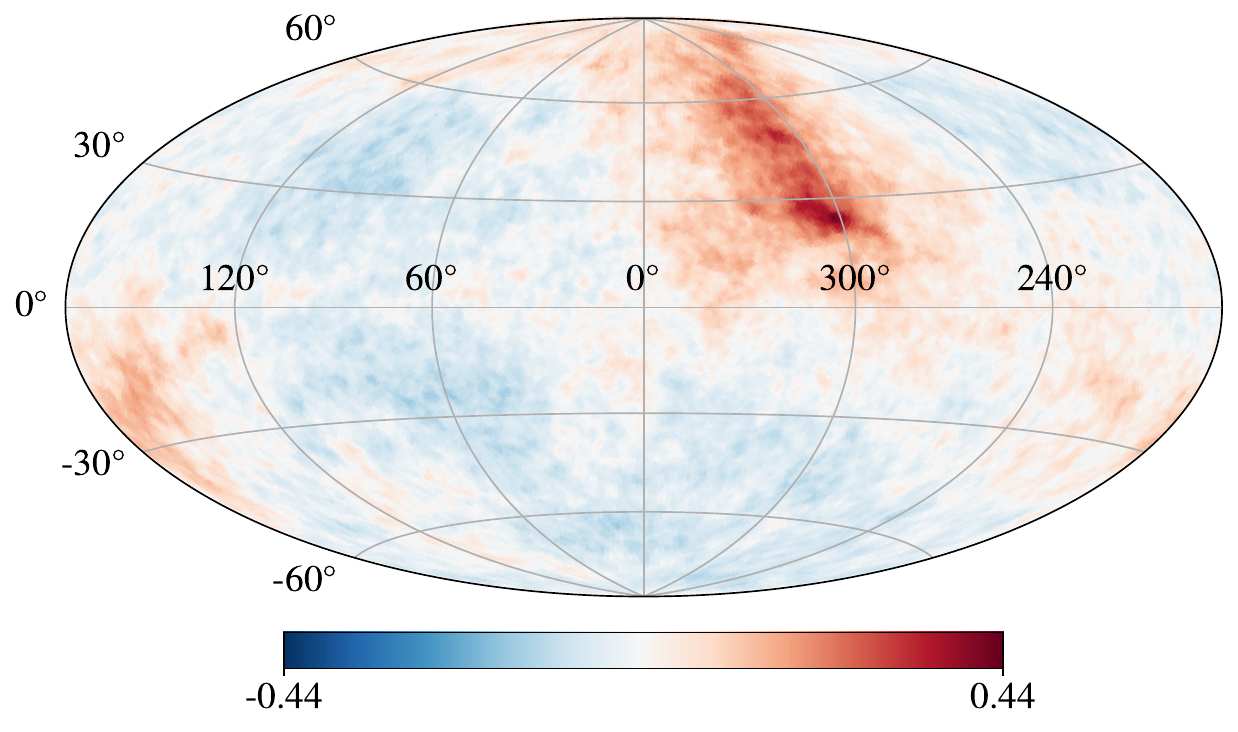}
    \includegraphics[width=\linewidth]{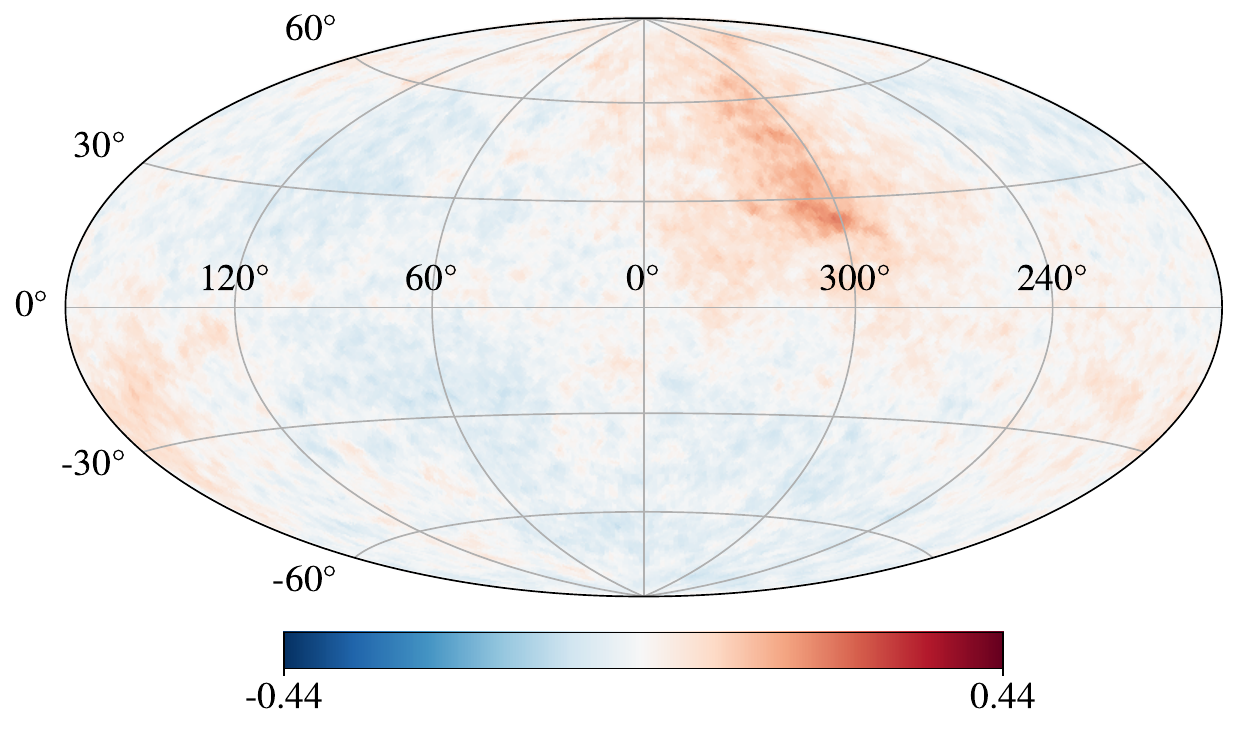}
    \caption{Predicted neutrino skymaps (in Galactic coordinates) with threshold energies at the maximum neutrino energy for two different source evolution indices: $m=-5$ (top) and $m=0$ (bottom). The color scale indicates the predicted neutrino flux relative to the all-sky average. We assume $z_0=2$ [roughly the peak of star-formation~\citep{Robertson:2015uda}] and $\gamma=-2.53$. The top skymap serves as the template map for our maximum likelihood analysis. For the reader's convenience, these maps can be found in equatorial coordinates in Appendix~\ref{app:raDec}.}
    \label{fig:skymaps}
\end{figure}

The predicted anisotropy can be seen in Fig.~\ref{fig:skymaps} for two different source evolutions: $m=-5, z_0=2$ (top) and $m=0, z_0=2$ (bottom). Both skymaps have an energy threshold at the cutoff energy $E_\mathrm{th} = E_\mathrm{max}$ and a source spectral index $\gamma=-2.53$, based on the IceCube Cascades dataset~\citep{IceCube:2020acn}. This anisotropy has a very particular structure owing to the imprint of the local LSS, which is maximized for small effective neutrino horizons (as in the $m=-5$ case) and is diminished as the contribution from high redshifts increases. In general, the neutrino skymap is roughly a sum of this maximal anisotropy and an isotropic contribution from higher redshift sources, whose relative contribution depends on the source evolution, the source spectrum, and the $E_\mathrm{th}/E_\mathrm{max}$ ratio. Therefore, we used a template analysis to characterize the level of anisotropy for a given neutrino skymap, rather than attempting to measure this anisotropy in terms of its multipole moments. We adopted the skymap in Fig.~\ref{fig:skymaps} (top) as our template for this purpose, since its parameters maximize this anisotropy by minimizing the effective neutrino horizon. Of course, this is not meant to be a realistic observational expectation for the neutrino sky, merely a tool to quantify the anisotropy level for more realistic scenarios. This template could be updated once the neutrino spectrum is more robustly measured at all energies, but we show the sensitivity of the inferred anisotropy level to a varying spectral index in the next section.

\par
We perform a maximum likelihood analysis to measure the anisotropy level of a skymap, $\alpha$, for a given source evolution compared to our template. We consider two nested hypotheses: an isotropic skymap as our null hypothesis and a superposition of an isotropic skymap and a local LSS anisotropy template as our alternative hypothesis. Under this framework the average number of neutrino events predicted in pixel $i$, $\mu_i$, is given by

\begin{align}
    \mu_i  = \left[\alpha \frac{J_{\nu,i}}{J_\nu} + (1-\alpha) \frac{\Omega_i}{4\pi}\right] N_\mathrm{evts}~,
\end{align}

\noindent
where $\alpha$ is the relative weight of the anisotropy template ($\alpha=0$ corresponding to the null hypothesis) and $N_\mathrm{evts}$ is the total number of neutrino events in the skymap. Given a set of observed neutrinos $n_i$, the likelihood is then given by the product of Poisson probabilities

\begin{align}
    \mathcal{L}(\alpha| n_i) = \prod_i \mathrm{Poiss}(n_i|\mu_i)~.
\end{align}

\noindent 
With this definition we obtain an estimate of $\alpha$ by maximizing the log-likelihood ratio $\ln\left(\mathcal{L}(\alpha|n_i)/\mathcal{L}(0|n_i)\right)$. In the plots below, unless otherwise specified, we report the maximum likelihood estimate (MLE) of $\alpha$ in the limit of large statistics, specifically assuming $N_\mathrm{evts}=10^6$.

\section{Results}
\par
Figure~\ref{fig:skymaps} illustrates the predicted neutrino anisotropy for two source evolutions: $m=-5$ and $m=0$. In both cases the energy threshold is at the cutoff energy to amplify the anisotropy from the local LSS. Specifically, for the assumed spectral index, evolution, and energy threshold the effective horizon is $z_\mathrm{eff} \simeq 0.29$ (about $1200$~Mpc comoving distance) and $z_\mathrm{eff} \simeq 0.65$ (about $2420$~Mpc comoving distance), respectively. These plots clearly demonstrate the isotropizing effect of having more sources at high redshifts (as is the case for more positive source evolutions).

\par
The most prominent structures in the skymaps can be attributed to structures in the local matter distribution. In particular, the anisotropy in the Northern Galactic sky is dominated by the Virgo and Coma superclusters along with the Great Attractor, while the features in the Southern Galactic sky are driven primarily by Perseus-Pisces supercluster~\citep{Bister:2023icg}.

\par
\begin{figure}
    \centering
    \includegraphics[width=\linewidth]{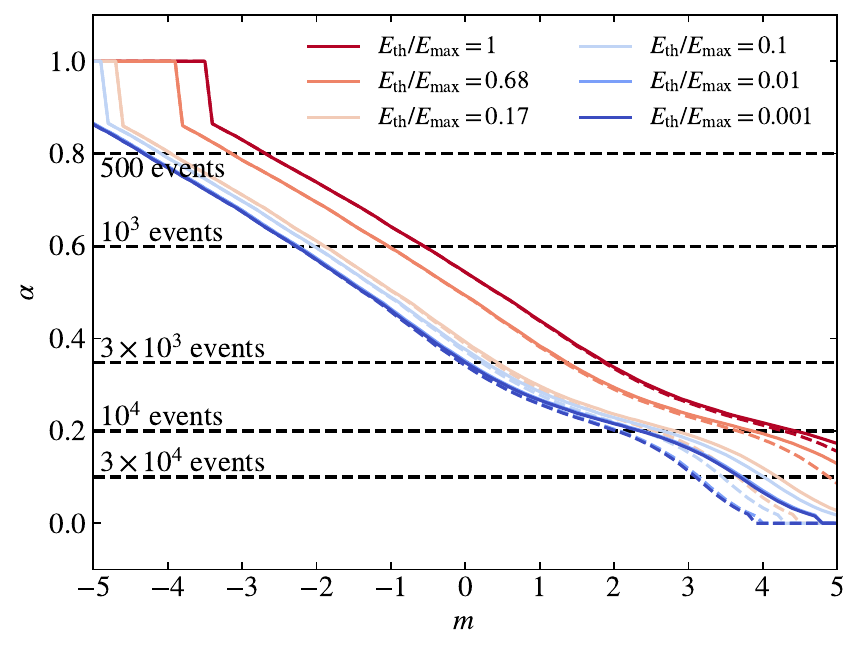}
    \caption{Anisotropy measure $\alpha$ as a function of source evolution power law index $m$ for various threshold energies (colored lines). All lines assume $\gamma=-2.53$, and $z_0=2$ (solid) or $z_0=4$ (dashed) to demonstrate the sensitivity to $z_0$. Dashed black lines indicate the $90\%$ confidence level upper-limit on $\alpha$ given a number of observed events for a truly isotropic distribution. Note that $\alpha$ saturates to $1$ $(0)$ once the level of anisotropy is maximized (minimized).}
    \label{fig:alpha_evo}
\end{figure}

The dependence of the predicted anisotropy level using our maximum likelihood analysis on source evolution is shown in Fig.~\ref{fig:alpha_evo}. As can be seen, there is a strong dependence of the anisotropy level on the source evolution. Since the other parameters in this plot can be measured directly (the neutrino spectral index and cutoff energy) or are set by the observer (the threshold energy) it is possible for observers to measure this quantity. Therefore, observers can use the level of anisotropy to measure the evolution of neutrino sources indirectly. Moreover, it is possible to place a lower bound on $m$ even if an observer can only place an upper-bound on $\alpha$. 

\par
From Fig.~\ref{fig:alpha_evo} it is clear there is a minimum level of anisotropy imprinted by the local LSS, which can be significant for negative source evolutions. This fact emphasizes that a measurement of the neutrino spectrum's cutoff energy is not necessary to perform this analysis. Fig.~\ref{fig:alpha_evo} also demonstrates that the anisotropy level can be enhanced by raising the threshold energy to shrink the observer-induced horizon, regardless of the source evolution.

\par
We note that one could already conduct such an analysis with $E_\mathrm{th} \lesssim 10$~PeV where IceCube currently measures a neutrino flux~\citep{IceCube:2023wmh,Naab:2023xcz}. The resulting anisotropy would enable measurement of the evolution of all neutrino sources contributing below UHEs, which are likely due to a mix of accelerators producing both sub-UHE CRs and UHECRs. Constraints on the evolution of neutrino sources would provide critical information for identifying which astrophysical source types produce the bulk of the astrophysical neutrino flux. In the future, if this analysis is done for $E_\mathrm{th} \gtrsim 30$~PeV, where neutrinos are only produced by UHECRs, this measurement would constrain or measure the evolution of UHECR sources alone. Such a constraint on $m$ would improve UHECR source modeling and be a step toward determining their sources. We present results for several specific UHECR-generated neutrino flux models in Appendix~\ref{app:modelSpecific}.

\par
\begin{figure}[htpb!]
    \centering
    \includegraphics[width=\linewidth]{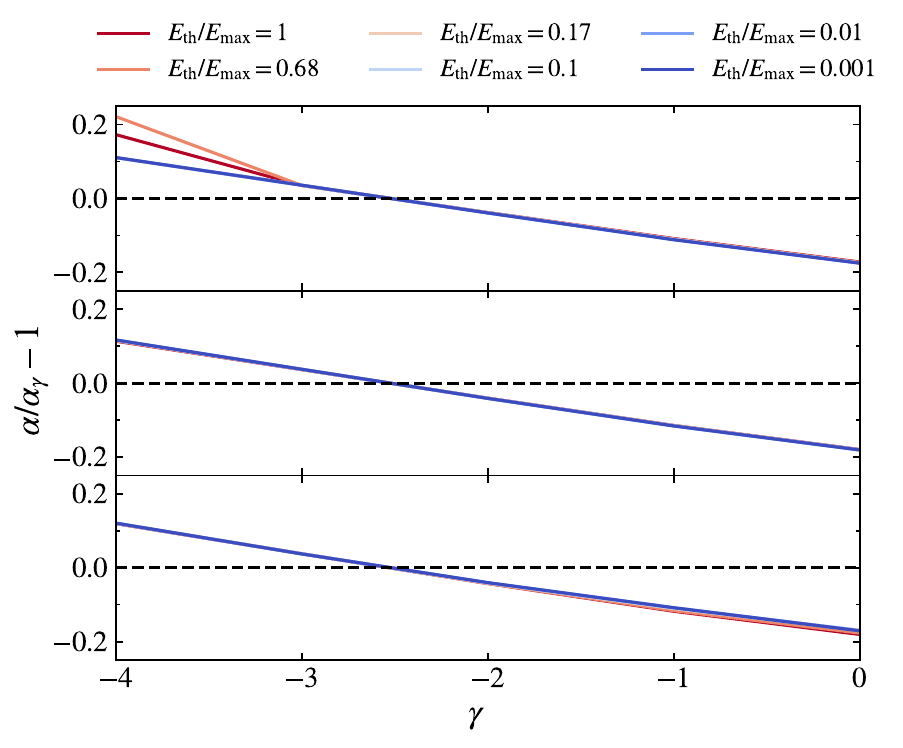}
    \caption{Relative difference of the inferred anisotropy level using a template with fixed spectral index, $\alpha$, compared to that obtained when using a template matching the true spectral index, $\alpha_\gamma$, as a function of the true spectral index. $\alpha$ is calculated with a template using $\gamma=-2.53$ and both templates assume $m=-5$. Results are shown for several true source evolutions ($z_0=2$ fixed): $m=-3$ (top), $m=0$ (middle), and $m=+3$ (bottom). Various $E_\mathrm{th}/E_\mathrm{max}$ ratios are shown as colored lines.}
    \label{fig:alpha_gamma}
\end{figure}

Observers will have a more accurate measurement of the anisotropy if the spectral index used to build the anisotropy template matches the true spectral index of the neutrino spectrum. To explore how the inferred anisotropy level is impacted by using a template whose spectral index does not match the true spectral index, we generate skymaps for various evolutions and spectral indices and apply our maximum likelihood analysis to them using two templates: one with a fixed spectral index $\gamma=-2.53$ and another with the true spectral index of the map. Both templates still assume an $m=-5$ evolution to represent the maximal level of anisotropy. Fig.~\ref{fig:alpha_gamma} shows the relative difference in the inferred level of anisotropy between these templates. In general, the level of anisotropy decreases for harder spectral indices, since harder spectral indices place a larger fraction of the source spectrum at high energies which effectively pushes the cutoff to slightly higher energies. This makes the soft horizon more gradual so that a larger fraction of the total flux comes from sources at high redshifts. As can be seen for true spectral indices from $\gamma=-3$ to $-2$ the error on the anisotropy is at the $5\%$~level or less. At the extreme, where the template's spectral index is significantly mismatched to the true spectral index, this error can be as large as roughly $25\%$. 

\par
The results discussed so far have assumed neutrinos are perfectly reconstructed and do not suffer from detector resolution effects. In the next section, we discuss the impact of these effects, as well as, other details which must be considered for such an analysis.

\section{Discussion}
\par
While the results discussed above represent a significant observational opportunity, they also represent a significant observational challenge. Positive source evolutions will have an inherently small neutrino anisotropy, making the exact source evolution difficult to measure. This situation can be improved slightly by raising the energy threshold to amplify the level of anisotropy, but this necessarily will decrease the number of neutrinos that can be used in such an analysis. In the following sections, we discuss the feasibility of detection (or constraint) of the neutrino anisotropy, details of setting the energy threshold, and the effects of energy resolution. Finally, we consider how the proposed analysis changes if we consider multiple neutrino source populations.

\subsection{Detectability of the predicted anisotropy}\label{sec:detectability}

\par
Here we investigate the detectability of the predicted anisotropy. To calculate this we use the log-likelihood ratio defined in Section~\ref{sec:model} as a test statistic (TS):

\begin{align}
    \mathrm{TS} = 2\ln\left(\frac{\mathcal{L}(\alpha|n_i)}{\mathcal{L}(0|n_i)}\right)~. 
\end{align}

\noindent
We calculate the TS distribution for an isotropic skymap, $\mathrm{TS}_0$, and a skymap with fixed anisotropy level $\alpha$, $\mathrm{TS}_\alpha$, for a given number of observed events $N$. The TS distributions allow us to find the minimal $N$ such that the cumulative tail probabilities satisfy $P(\mathrm{TS}_0\geq \mathrm{TS}_ \mathrm{crit}) = \beta$ and $P(\mathrm{TS}_\alpha \leq \mathrm{TS}_ \mathrm{crit}) = \epsilon$, where $1-\beta$ corresponds to the significance with which the null hypothesis is rejected at $1-\epsilon$ confidence level. 

\par
\begin{figure}
    \centering
    \includegraphics[width=\linewidth]{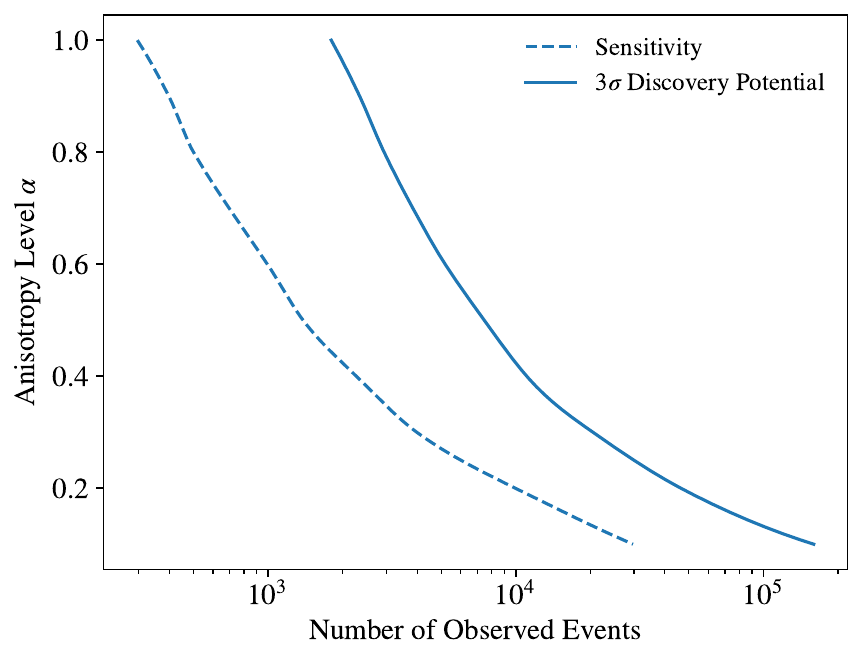}
    \caption{The minimal anisotropy level $\alpha$ to which a detector is sensitive (dashed line) or has the potential to measure with $3\sigma$ significance (solid line) as a function of the number of observed neutrino events.}
    \label{fig:detectability}
\end{figure}

Following the convention of high-energy neutrino astronomy~(see e.g.~\citet{IceCube:2025zyb}), we define a detector's \textit{sensitivity} to an anisotropy level $\alpha$ to be the minimal $N$ such that $90\%$ of the $\mathrm{TS}_\alpha$ distribution is greater than the median of the $\mathrm{TS}_0$ distribution (i.e.\ $\beta = 0.5$, $\epsilon = 0.1$). On the other hand, a detector's \textit{$3\sigma$ discovery potential}\footnote{We are using here a terminology common in the field, ``discovery potential", which refers to the level at which a detector could (at a fixed confidence level) measure a signal with a particular statistical significance.} is the minimal $N$ such that $50\%$ of the $\mathrm{TS}_\alpha$ distribution is above $3\sigma$ of the $\mathrm{TS}_0$ distribution (i.e.\ $\beta \simeq 1.35\times 10^{-3}$, $\epsilon = 0.5$). Figure~\ref{fig:detectability} shows the minimal value of $\alpha$ to which a detector with $N$ observed events is sensitive or has the potential to measure with $3\sigma$ significance. We note that the sensitivity also represents the constraint that can be placed on $\alpha$ in the event of a non-detection.

\subsection{Choosing an energy threshold}\label{sec:energyThresh}

\par
A number of factors will contribute to the choice of threshold energy for observers carrying out the proposed analysis. These include statistical significance, the underlying level of anisotropy, and the target population whose evolution is being investigated. 

\par
The first of these factors is likely the most important: observers must have enough extragalactic neutrino events in their above-threshold dataset to achieve sensitivity to the predicted anisotropy. This means that $>300$~astrophysical neutrino events are needed to carry out this analysis. Larger neutrino datasets are available for lower energy thresholds, but low energies may lose some sensitivity due to contamination from the Galactic plane foreground and atmospheric neutrinos.

\par
To give a sense of the range of possible extragalactic neutrino events in the IceCube dataset, as well as, projected for IceCube-Gen2 for various energy thresholds we explore two extreme flux models. As a minimal flux model, we adopt the single-power law fit to the astrophysical neutrino spectrum obtained by~\citet{Naab:2023xcz} from $10$~TeV to $6.3$~PeV. Beyond $6.3$~PeV we assume that neutrino flux is sharply cutoff by a step function. As a maximal flux model, we adopt the same single-power law fit as in the minimal model, but assume that beyond $6.3$~PeV the true flux saturates the global $90\%$ CL upper-bound from IceCube~\citep{IceCube:2025ezc} and ANITA~\citep{ANITA:2019wyx}. These two models represent the extreme models for the astrophysical neutrino flux allowed by current constraints. In both cases, we assume a $1:1:1$~flavor ratio.

\par
For each of these models, we estimate the expected number of astrophysical events for $10$~years of livetime at analysis-level for IceCube and IceCube-Gen2.\footnote{We note that IceCube has not published an estimate of the total number of astrophysical neutrinos in its dataset.} For IceCube, this is calculated by combining the effective areas of the Northern Tracks~\citep{Abbasi:2021qfz,Naab:2023xcz,Muzio:2025gbr} and DNN Cascades~\citep{IceCube:2023ame} datasets below $10$~PeV with the latest EHE effective area~\citep{IceCube:2025ezc} at higher energies. For IceCube-Gen2 we combine the effective area of the optical array to track-like events with the effective area of the radio array~\citep{IceCubeGen2_TDR}. The resulting number of events for various energy thresholds are presented in Table~\ref{tab:expEvents}. We note that these events do not necessarily sample all parts of the sky equally.

\begin{table}[h!]
    \centering
    \begin{tabular}{c|c|c}
        \hline
        \textbf{$\lg{E_\mathrm{th}/\mathrm{eV}}$} & IceCube & IceCube-Gen2 \\
        \hline
        $13.0$ & $4253$ ($4241$) & $11559$ ($11118$) \\
        $14.0$ & $612$ ($600$) & $2347$ ($1905$) \\
        $15.0$ & $52$ ($41$) & $610$ ($168$) \\
        $16.0$ & $8$ ($0$) & $415$ ($0$) \\
        $17.0$ & $5$ ($0$) & $380$ ($0$) \\
        $18.0$ & $3$ ($0$) & $328$ ($0$) \\
        $19.0$ & $1$ ($0$) & $189$ ($0$) \\
        $19.5$ & $0$ ($0$) & $88$ ($0$) \\
        \hline

    \end{tabular}
    \caption{Estimated total number of expected astrophysical events for the maximal (minimal) flux model above various energy thresholds after $10$~years of livetime at analysis-level in IceCube and IceCube-Gen2. See text for model details.}\label{tab:expEvents}
\end{table}

\par
Due to the falling spectrum, the number of neutrino events that would be realistically available to search for an anisotropy strongly depends on the energy threshold chosen, as Table~\ref{tab:expEvents} shows. The current IceCube dataset could in principle have sufficient statistics to measure anisotropy levels as low as $\alpha\simeq 0.7$ and constrain anisotropy levels down to $\alpha\simeq 0.3$ with $E_\mathrm{th} = 10^{13}$~eV. Comparing to Fig.~\ref{fig:alpha_evo}, this means that IceCube could potentially measure source evolutions $m\lesssim -3$ and exclude all source evolutions more negative than $m\simeq +0.5$. Slightly higher thresholds could also be used at the cost of reduced sensitivity to low anisotropy levels. IceCube-Gen2, on the other hand, would have a large enough exposure to constrain anisotropy levels as low as $\alpha \simeq 0.2$ and the potential to measure anisotropy levels $\alpha \gtrsim 0.4$ with $E_\mathrm{th}= 10^{13}$~eV. Again comparing with Fig.~\ref{fig:alpha_evo}, this means that IceCube-Gen2 could exclude source evolutions more negative than $m\simeq 2$ or measure source evolutions more negative than $m\simeq -0.5$.

\par
Finally, the energy threshold chosen will determine the source populations whose evolution is probed by the analysis. At energies above ${\sim}30$~PeV, neutrinos are produced only by UHECRs. Therefore, analyses with energy thresholds $\gtrsim30$~PeV will probe the evolution of UHECR sources alone. At lower energies, both lower-energy sources and UHECRs (particularly via via hadronic interactions in their sources) can contribute to the observed neutrino flux. Analyses with energy thresholds below $30$~PeV will therefore probe the evolution of sources which most significantly contribute to the observed astrophysical neutrino flux. A more detailed discussion of the applicability of this analysis to a neutrino flux produced by multiple source populations can be found in Section~\ref{sec:nonStdSrcs}.

\subsection{Effect of neutrino detector energy resolution}\label{sec:energyRes}

\par
Throughout the main text we have assumed that the energy of observed neutrinos is precisely known. However, astrophysical neutrino detectors have large uncertainties on the neutrino energy. For example, for the IceCube detector the estimated energy resolution ranges from $11\%$ for cascade-type events to $30\%$ for starting track-type events~\citep{IceCube:2020wum}. This has the effect of further blurring the observer-induced neutrino horizon. 

\par
Since the neutrino spectrum is falling, energy resolution effects will lead to a migration of lower energy neutrino events into higher energy bins in observation. This has two effects: slightly lowering the true energy threshold from the intended threshold, while also slightly hardening the observed spectrum. From Fig.~\ref{fig:alpha_evo} and Fig.~\ref{fig:alpha_gamma} we expect that both of these effects will slightly diminish the expected anisotropy level. 

\par
To understand the approximate size of this effect we adopt a Gaussian blurring $\kappa(E, E_\mathrm{true}; f_\sigma)$ of the true neutrino energy given by

\begin{align}
    \kappa(E, E_\mathrm{true}; f_\sigma) = \frac{1}{\sqrt{2\pi}f_\sigma E_\mathrm{true}} \exp\left( - \frac{(E-E_\mathrm{true})^2} {2(f_\sigma E_\mathrm{true})^2} \right)~,
\end{align}

\noindent
where $E$ is the observed neutrino energy, $E_\mathrm{true}$ is the true neutrino energy, and $f_\sigma = \sigma(E_\mathrm{true})/E_\mathrm{true}$ is the detector energy resolution.\footnote{In reality, the detector resolution is non-Gaussian with a long tail towards underestimating the true neutrino energy~\citep{IceCube:2020wum}. However, we leave a more detailed study of detector resolution effects to future work.} With this blurring $Q_\nu(E)$ in equation~\eqref{eq:totalFlux} is given by

\begin{align}
    Q_\nu(E) = \int_0^\infty Q(E_\mathrm{true}) \kappa(E, E_\mathrm{true}; f_\sigma) dE_\mathrm{true}~.
\end{align}

\par
\begin{figure}
    \centering
    \includegraphics[width=\linewidth]{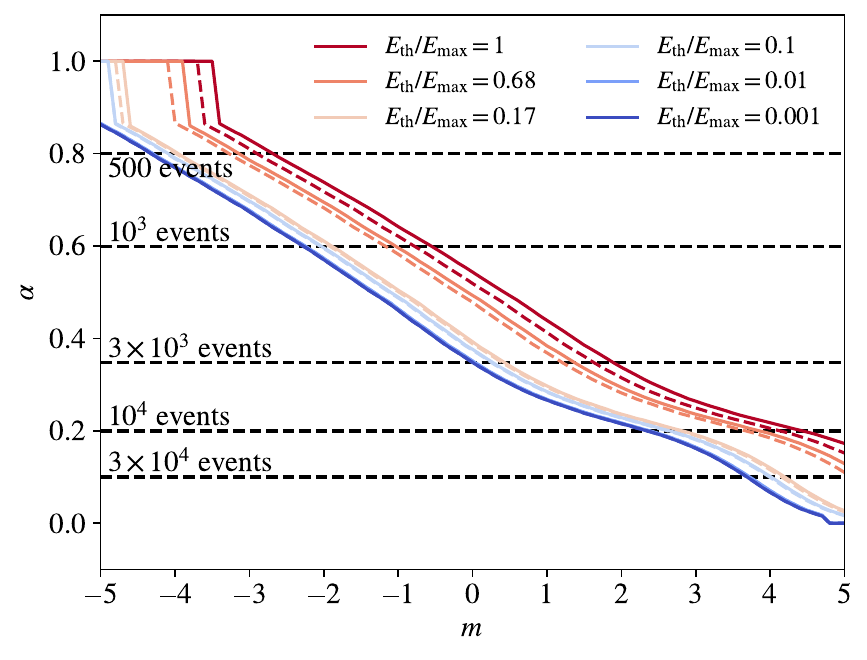}
    \caption{Anisotropy measure $\alpha$ as a function of source evolution power law index $m$ for various threshold energies (colored lines) with perfect (solid) and $30\%$ (dashed) energy resolutions. All lines assume $\gamma=-2.53$ and $z_0=2$. Dashed black lines same as in Fig.~\ref{fig:alpha_evo}.}
    \label{fig:alpha_evo_energyRes}
\end{figure}

In Fig.~\ref{fig:alpha_evo_energyRes} we show the predicted anisotropy level for different energy thresholds with perfect energy resolution compared to a $30\%$ energy resolution. The level of anisotropy is moderately diminished, as expected, for energy thresholds near the maximum neutrino energy. For energy thresholds far below the cutoff energy, energy resolution effects are irrelevant since they do not impact the effective neutrino horizon.

\subsection{Effect of cutoff energy variance of sources}\label{sec:EmaxVariance}

\par
So far we have assumed that a single population of identical sources produces neutrinos. In this Section, we consider the impact of a neutrino source population whose cutoff energy follows a distribution on the expected neutrino anisotropy. For simplicity, we assume these sources are otherwise of a single type, with a common spectral index, and have a common source evolution; these assumptions are separately explored in Section~\ref{sec:nonStdSrcs}. 

\par
To investigate the impact of population variance in the cutoff energy of the emitted neutrino spectrum, we follow the methodology of~\citet{Ehlert:2022jmy}. In particular, we assume that the distribution of maximum energies for individual sources, $E_\mathrm{max,src}$, follows a Pareto distribution:

\begin{align}
    p(E_\mathrm{max,src}) = 
    \begin{cases} 
      0 & E_\mathrm{max,src} < E_\mathrm{max}  \\
      \frac{\beta_\mathrm{pop}-1}{E_\mathrm{max,src}} \left( \frac{E_\mathrm{max,src}}{E_\mathrm{max}} \right)^{-\beta_\mathrm{pop}} & E_\mathrm{max,src} \geq E_\mathrm{max}
    \end{cases}~,
\end{align}

\noindent
where $\beta_\mathrm{pop}$ is the power-law index of the distribution above the minimum population cutoff energy, $E_\mathrm{max}$. Sources with smaller values of $\beta_\mathrm{pop}$ have a larger variance of maximum energies. As shown in~\citet{Ehlert:2022jmy}, this distribution of maximum energies (for a population of sources producing single power-law spectra with exponential cutoff) results in a population spectrum of the form

\begin{align}
    Q_{\nu,\mathrm{pop}} \propto E^\gamma \left(\frac{E}{E_\mathrm{max}}\right)^{1-\beta_\mathrm{pop}} \gamma_\mathrm{inc}\left(\beta_\mathrm{pop}-1, \frac{E}{E_\mathrm{max}}\right)~,
\end{align}

\noindent
where $\gamma_\mathrm{inc}$ is the lower incomplete gamma function. As can be seen from Fig.~1 of~\citet{Ehlert:2022jmy}, the population distribution primarily serves to broaden the spectral cutoff, which is roughly equivalent to raising the cutoff energy. We can then assess the impact of such a population variance on the predicted anisotropy simply by replacing $Q_\nu$ by the population spectrum $Q_{\nu,\mathrm{pop}}$ in equation~\eqref{eq:totalFlux}. 

\begin{figure}
    \centering
    \includegraphics[width=\linewidth]{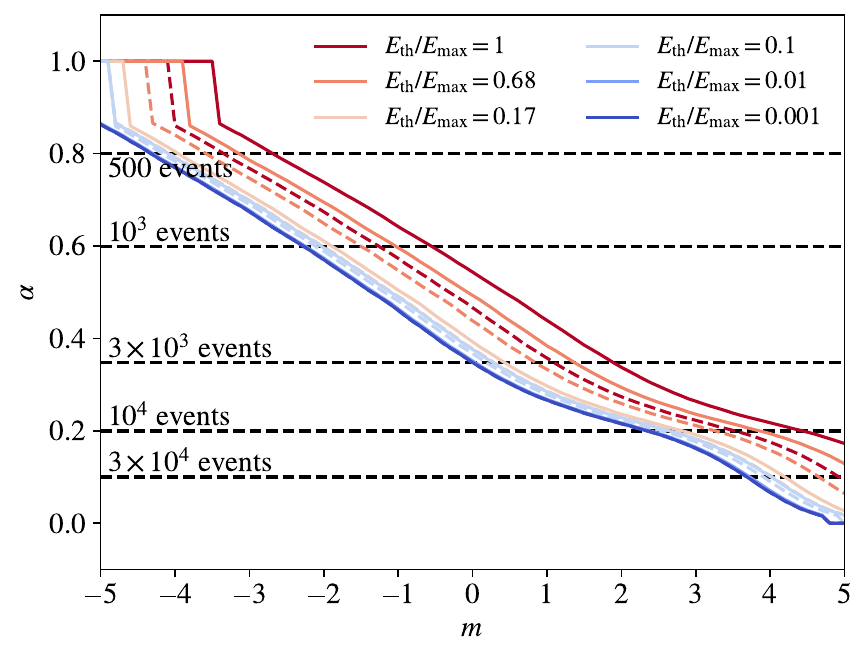}
    \caption{Anisotropy measure $\alpha$ as a function of source evolution power law index $m$ for various threshold energies (colored lines) for sources with identical maximum energies (solid) and a distribution of maximum energies with $\beta_\mathrm{pop} = 3.0$ (dashed). All lines assume $\gamma=-2.53$ and $z_0=2$. Dashed black lines same as in Fig.~\ref{fig:alpha_evo}.}
    \label{fig:alpha_evo_popVariance}
\end{figure}

\par
Figure~\ref{fig:alpha_evo_popVariance} shows the impact of a population with the variance in $E_\mathrm{max}$. Specifically, we adopt the smallest best-fit value of $\beta_\mathrm{pop}$ minus $1\sigma$ (i.e.\ $\beta_\mathrm{pop}=3.0$) of any scenario presented in~\citet{Ehlert:2022jmy}. This corresponds to a population with $90\%$ of sources producing maximum energies, $E_\mathrm{max,src}$, within a factor of about $3$ of the minimum population cutoff energy, $E_\mathrm{max}$. Since the population spectrum's effective cutoff energy is higher than $E_\mathrm{max}$, the effect of such a population variance is equivalent to lowering the ratio $E_\mathrm{th}/E_\mathrm{max}$ in case of perfectly identical neutrino sources. In particular, far below $E_\mathrm{max}$ such a population variance is irrelevant since it does not impact the effective neutrino horizon.

\subsection{Application to multiple source populations}\label{sec:nonStdSrcs}

\par
An important simplifying assumption made in Section~\ref{sec:model} was that all sources emit a common neutrino spectrum -- in other words, all sources are assumed to be standard (up to their luminosity-density, which is parametrized by the source evolution). Here we explore the applicability of the analysis proposed in Section~\ref{sec:model} in the case that sources are non-standard and that several source populations contribute significantly to the observed neutrino flux. We consider two cases: that different source types have different spectral properties but follow the same source evolution; and, the more general case, that different source types have both different spectral properties and evolutions.

\par
In the former case, the observed neutrino flux is a sum of fluxes produced by each source population. Since we assume that these sources have a common evolution, this observed flux can be obtained from equation~\eqref{eq:totalFlux} by simply defining $Q_\nu(E)$ as a sum over source populations $i$,

\begin{align}
    Q_\nu(E) = \sum_i Q_{\nu,i}(E)~.
\end{align}

\noindent
In this case, the analysis proposed can be carried out in the same way as for a single, standard source population.

\par
In the case that their are multiple source populations with different spectral properties and evolutions, a slightly modified version of the analysis can be used. In this case, the product $\mathcal{H}(z)Q_\nu(E)$ in equation~\eqref{eq:totalFlux} can be written as a sum over source populations $i$,

\begin{align}
    \mathcal{H}(z)Q_\nu(E) = \sum_i \mathcal{H}_i(z)Q_{\nu,i}(E)~.
\end{align}

\noindent
However, since the source evolution cannot be factored out of this sum, our analysis would require each population's source evolution to be modeled separately. In this case, the constraining power of anisotropy data on each of these source evolutions would be significantly reduced. However, one can use this analysis to constrain an effective source evolution $\mathcal{H}_\mathrm{eff}$,

\begin{align}
    \mathcal{H}_\mathrm{eff}(z; E_\mathrm{th}) = \sum_i \mathcal{H}_i(z) \frac{\int_{E_\mathrm{th}}^\infty Q_{\nu,i}((1+z)E) dE}{\sum_j \int_{E_\mathrm{th}}^\infty Q_{\nu,j}((1+z)E) dE}~,
\end{align}

\noindent
a flux-weighted average of the evolutions of the underlying source populations. In this case $\mathcal{H}_\mathrm{eff}$ is a function of $z$ and $E_\mathrm{th}$. With this definition, we can rewrite the integral $\int_{E_\mathrm{th}}^\infty \mathcal{H}(z)Q_\nu((1+z)E) dE$ in equation~\eqref{eq:totalFlux} as

\begin{align}
    \int_{E_\mathrm{th}}^\infty \mathcal{H}(z)&Q_\nu((1+z)E) dE \nonumber \\ 
    &= \mathcal{H}_\mathrm{eff}(z, E_\mathrm{th}) \sum_i \int_{E_\mathrm{th}}^\infty Q_{\nu,i}((1+z)E) dE~.
\end{align}

\noindent
Anisotropy data will then constrain $\mathcal{H}_\mathrm{eff}$, rather than each source population's evolution individually. Moreover, by changing the threshold energy anisotropy data would also be able to probe the evolution of the dominant source populations within the horizon set by $E_\mathrm{th}$. In this way, the procedure laid out in Section~\ref{sec:model} can still be applied even in the case of non-standard source populations.

\subsection{Other factors impacting the neutrino anisotropy}\label{sec:otherFactors}

\par
Throughout this \paper{} we have assumed the observed neutrino flux is dominated by neutrinos of astrophysical origin. The flux of cosmogenic neutrinos depends on the details of the UHECR spectrum escaping the source. However, the majority of the cosmogenic neutrino flux arriving at Earth is produced by UHECRs originating from sources at higher redshifts and will therefore inherit the isotropy of high redshift sources. Thus, a cosmogenic contribution to the observed neutrino flux will mostly serve to reduce the level of anisotropy but a specific model of the UHECR flux escaping sources is required to quantify by how much. It is worth noting, however, that some energy regimes are more likely to be dominated by astrophysical neutrinos than others. In particular, UHECR source modeling~(e.g.~\citet{Muzio:2021zud}) shows that neutrinos in the $1-100$~PeV range are likely to be dominantly produced inside sources, while those $\gtrsim1$~EeV are likely to be equally, or dominantly in some cases, of cosmogenic origin~(e.g.~\citet{2017ApJ...839L..22G}). However, the evolution of $\gtrsim$~EeV neutrino sources can be constrained using other methods~\citep{Muzio:2023skc} which strongly complement those discussed in this work --- namely, the methods discussed in~\citet{Muzio:2023skc} are most sensitive in the cosmogenically-dominated case while those discussed here are most sensitive in the astrophysically-dominated case. In any case, due to the falling neutrino spectrum, the integrated neutrino flux will always be dominated by lower energy events so it is unlikely for a cosmogenic flux to have a significant impact on the level of anisotropy unless the energy threshold is near the EeV-scale.

\par 
So far in this discussion, we have ignored the effects of extragalactic magnetic fields. Strong extragalactic magnetic fields around filaments in the LSS may increase the residence time of UHECRs in their source clusters~\citep{Kotera:2008ae,Kotera:2009ms,Harari:2016vtz,Fang:2017zjf,Condorelli:2023xkx}. This would serve to produce an additional neutrino flux contribution which traces the matter distribution of the universe, thereby further increasing the neutrino anisotropy. However, this effect is strongly model-dependent and is beyond the scope of this study. 

\par
Finally, we note that the bias between the normal matter and dark matter density fields may affect the exact level of neutrino anisotropy. This bias can depend on the properties of the galaxy host to the neutrino source, which may lead some source types to produce neutrino anisotropies that are not well-represented by the simple model described above. Additionally, other beyond the Standard Model processes may generate UHE neutrinos that are isotropic or have a different anisotropy than that predicted. We leave the investigation of these effects for future work, but note that our predictions here clarify the Standard Model-only expectation.

\section{Summary}

\par
In this \paper{} we investigated the anisotropy of the UHE neutrino sky due to the local LSS. We found that the level of anisotropy can be used to constrain the evolution of neutrino sources and demonstrated how observers can enhance this anisotropy by selecting an energy threshold. In particular, if these neutrinos are sufficiently high energy, such a measurement would place strong constraints on the evolution of UHECR sources and provide a critical probe of their origin. 

\par
To detect or constrain this anisotropy, we showed that a large neutrino data set is required. The existing IceCube dataset is likely large enough to measure anisotropies larger than $\alpha \simeq 0.7$ and to constrain the anisotropy level down to $\alpha \simeq 0.3$. This means that IceCube can already measure neutrino source evolutions more negative than $m\simeq -3$ and exclude source evolving with $m\lesssim +0.5$. IceCube-Gen2~\citep{IceCube-Gen2:2020qha} will have sufficient exposure to measure anisotropies as low as $\alpha\simeq 0.4$, and therefore evolutions more negative than $m\simeq-0.5$, or place constraints on the anisotropy down to $\alpha\simeq 0.2$, excluding evolutions $m\leq 2$. Additional sensitivity, especially at higher energies, could be gained through a coordinated effort among the next generation of neutrino observatories, such as IceCube-Gen2 and GRAND~\citep{GRAND:2018iaj}. The resulting constraints or measurements would be an invaluable data point in the modeling of both neutrino and UHECR sources.

\par
Beyond measurement of the source evolution itself, observation of the neutrino anisotropy would represent a breakthrough in overcoming confounding factors in UHECR studies. In particular, it would provide an independent confirmation that the origin of the UHECR dipole is linked to the LSS. Moreover, this would itself provide an additional constraint on models of the Galactic magnetic field, which predict significantly different UHECR deflections but fit observational data equally well~(e.g.~\citet{Unger:2023zbg}).\\

\begin{acknowledgments}

We thank Yehuda Hoffman for his permission to use the
density field from~\citet{Hoffman:2018ksb}. We thank Brian Clark, Lu Lu, William Luszczak, Michael Unger, Stephanie Wissel, and Tianlu Yuan for helpful feedback and suggestions. The research of M.S.M.\ is supported by the NSF MPS-Ascend Postdoctoral Award \#2138121 and the John Bahcall Fellowship at the Wisconsin IceCube Particle Astrophysics Center at the University of Wisconsin--Madison. This work was supported by a grant from the Simons Foundation (00001470, NG).

\end{acknowledgments}

\appendix

\section{Neutrino skymaps in equatorial coordinates}\label{app:raDec}

\par
\begin{figure}[htpb!]
    \centering
    \includegraphics[width=\linewidth]{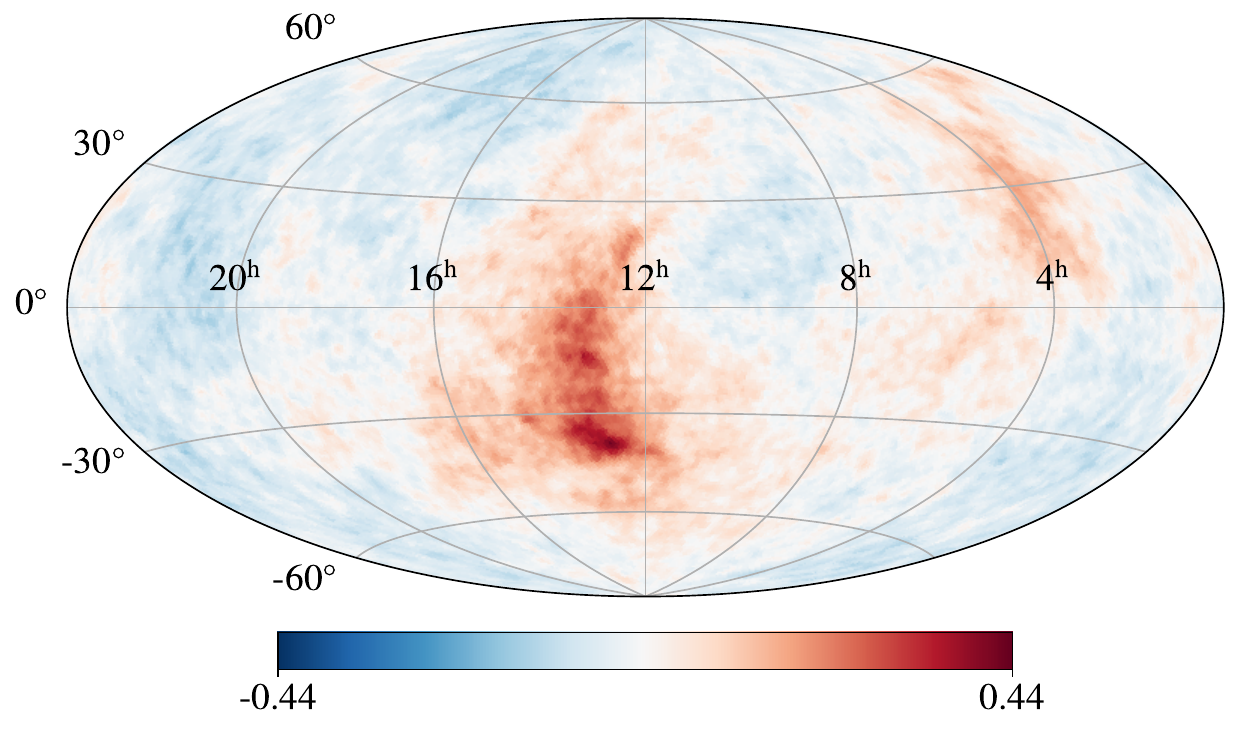}
    \includegraphics[width=\linewidth]{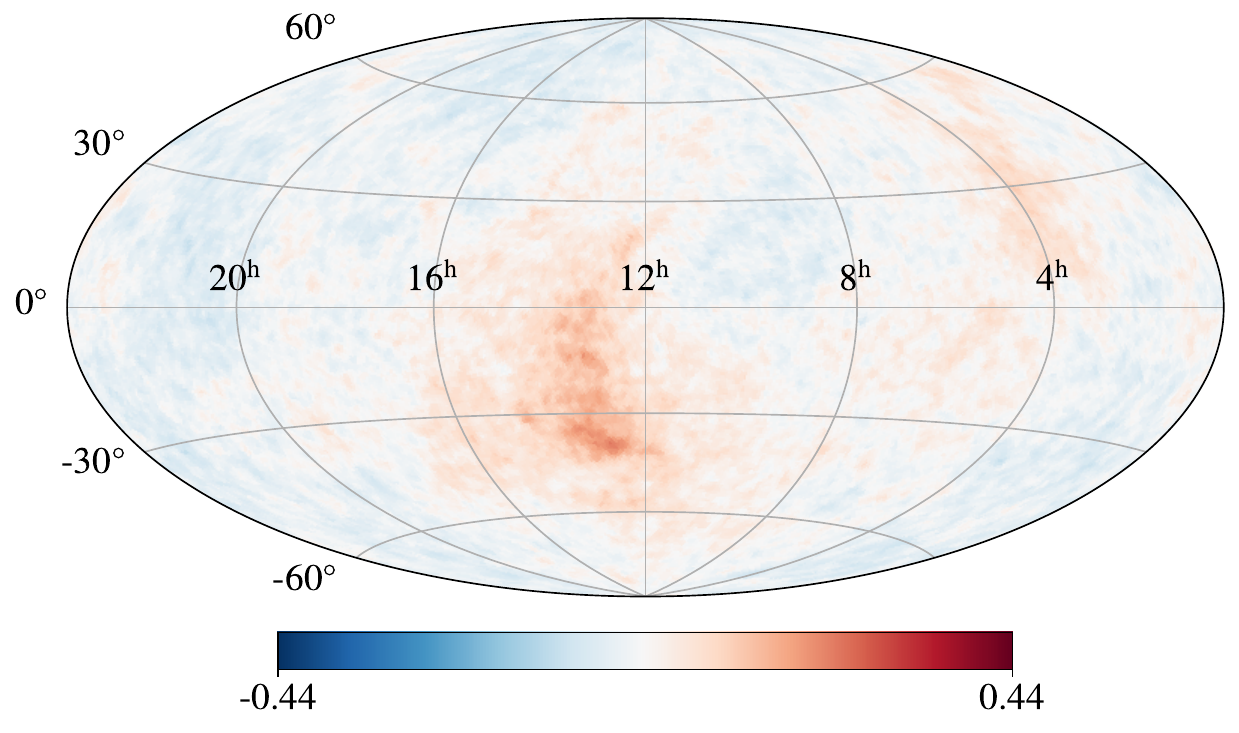}
    \caption{Same as in Fig.~\ref{fig:skymaps} but in equatorial coordinates.}
    \label{fig:skymaps_eq}
\end{figure}

Figure~\ref{fig:skymaps_eq} shows the predicted neutrino skymaps in equatorial coordinates with threshold energy at the maximum neutrino energy for two different source evolution indices: $m = -5$ (top) and $m = 0$ (bottom). These are the same skymaps as shown in Fig.~\ref{fig:skymaps}.

\section{Model-specific skymap examples}\label{app:modelSpecific}

\par
Here we apply our methodology to three specific UHE neutrino flux models, rather than the exponentially cutoff single power-law spectrum used for illustration in the main text. For all the examples below we take a low threshold energy of $E_\mathrm{th} = 10^{14}$~eV for the skymaps. 

\par
\begin{figure}[htpb!]
    \centering
    \includegraphics[width=\linewidth]{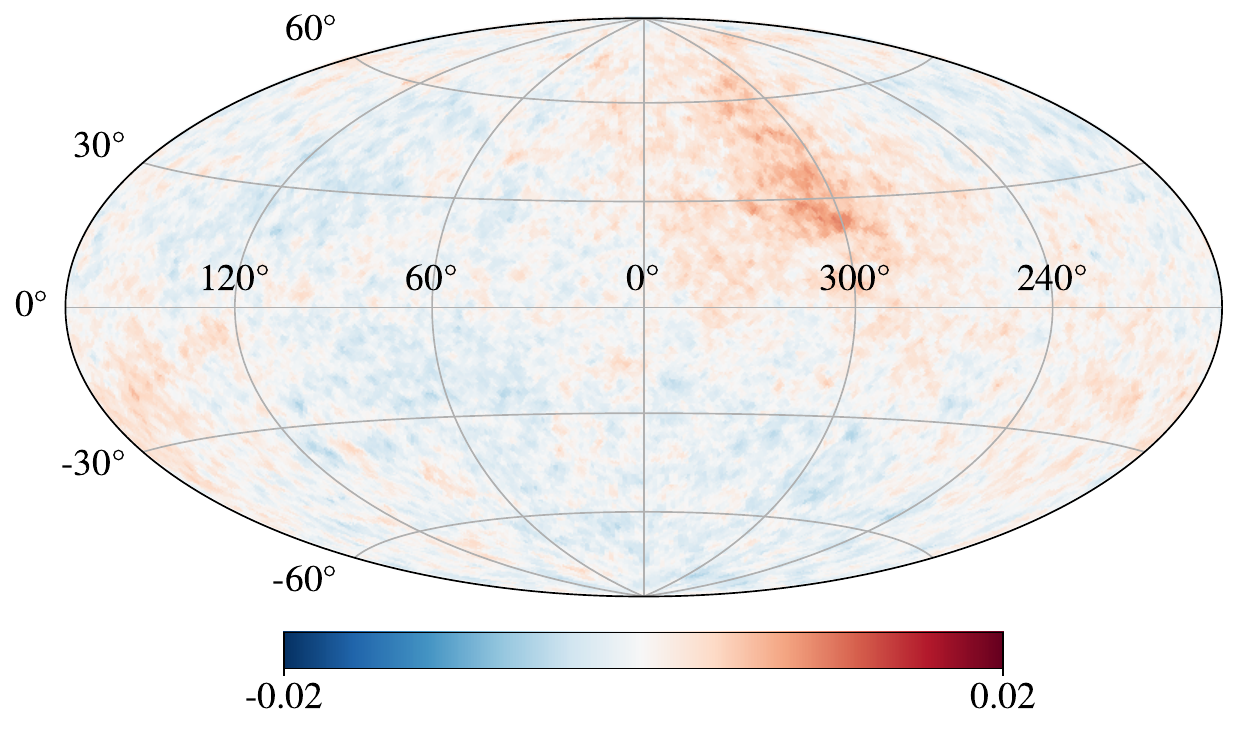}
    \caption{The predicted neutrino skymap (in Galactic coordinates) for source neutrinos predicted in Model A of \citet{Globus:2014fka}. This map was produced integrating above $E_\mathrm{th} = 10^{14}$ eV.}
    \label{fig:skymap_Globus}
\end{figure}

As a first example, Fig.~\ref{fig:skymap_Globus} shows the predicted neutrino skymap for the neutrino flux model from \citet{Globus:2014fka} (see Fig.~35 model A therein). Here we only consider the neutrino flux produced by UHECR interactions inside of GRB internal shocks. We adopt the GRB source evolution given by equation~28 of \citet{Globus:2014fka}. 

\par
We applied the MLE described in Section~\ref{sec:model} to determine the level of anisotropy, using the same anisotropy template as in the main text. This resulted in an anisotropy level of $\alpha\simeq 0.007$. This low level of anisotropy is consistent with the results shown in Figs.~\ref{fig:alpha_evo} and~\ref{fig:alpha_gamma}, mostly driven by the positive source evolution of GRBs ($m\simeq2$) which is further suppressed due to hard predicted neutrino spectrum.

\par
\begin{figure}[htpb!]
    \centering
    \includegraphics[width=\linewidth]{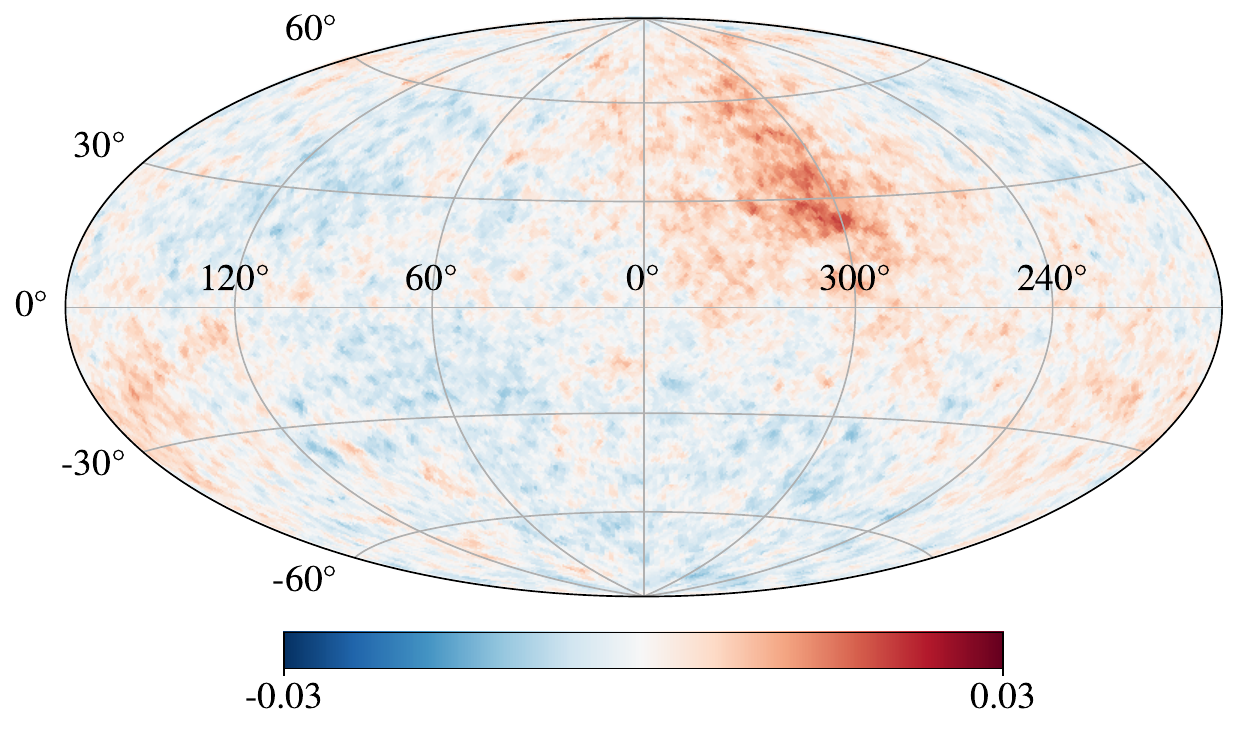}
    \caption{The predicted neutrino skymap (in Galactic coordinates) for the neutrino flux predicted in \citet{Fang:2017zjf}. This map was produced integrating above $E_\mathrm{th} = 10^{14}$ eV.}
    \label{fig:skymap_FangMurase}
\end{figure}

As a second example, we consider the neutrino flux predicted in~\citet{Fang:2017zjf} from UHECRs produced in AGN jets interacting inside the host cluster. We adopt the source evolution parameters of this model ($m=3$, $z_0=1.5$). The resulting skymap is shown in Fig.~\ref{fig:skymap_FangMurase}.

\par
Above roughly $0.2$~EeV this flux is dominated by cosmogenic neutrinos, which will tend contribute isotropically to the neutrino sky. However neutrinos above $0.2$~EeV contribute roughly $0.01\%$ of the total integrated flux, so we neglect this contribution to the skymap. This results in an anisotropy level of $\alpha\simeq 0.074$ using our anisotropy template. Again, this level of anisotropy is consistent with our results in the main text.

\par
\begin{figure}[htpb!]
    \centering
    \includegraphics[width=\linewidth]{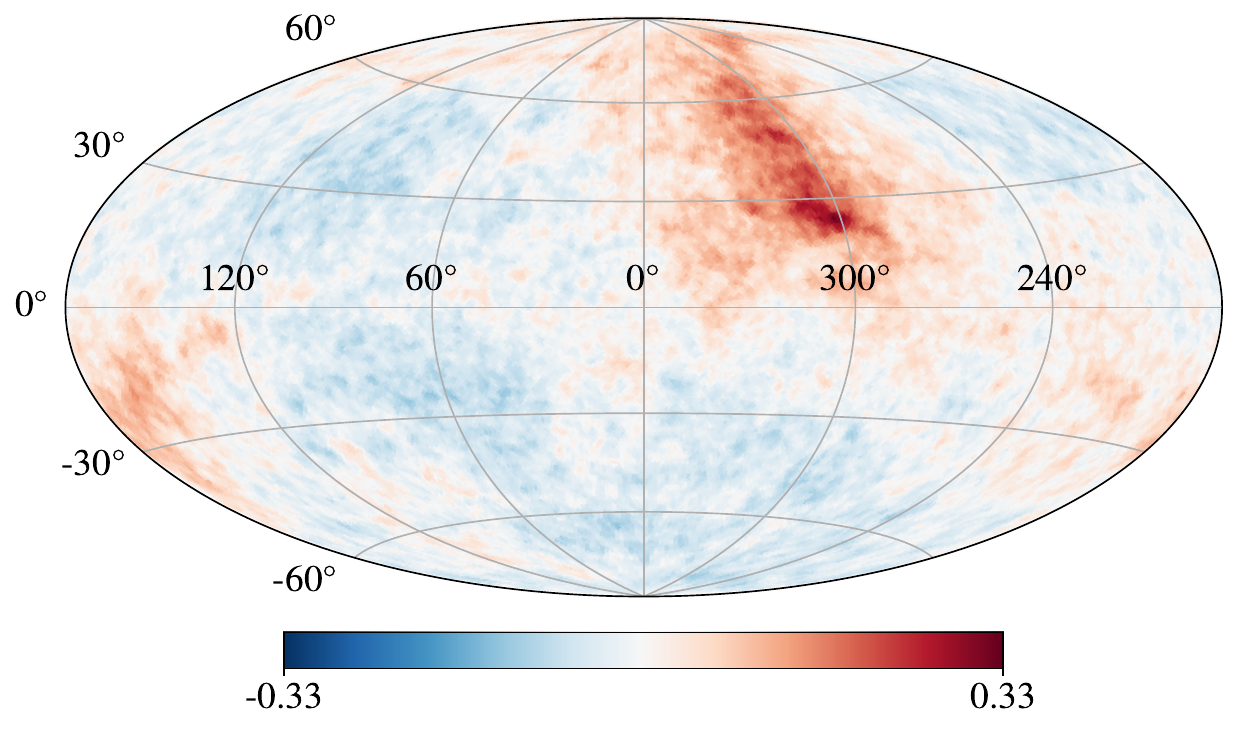}
    \caption{The predicted neutrino skymap (in Galactic coordinates) for the neutrino flux predicted in \citet{Rodrigues:2020pli}. This map was produced integrating above $E_\mathrm{th} = 10^{14}$ eV.}
    \label{fig:skymap_Rodrigues}
\end{figure}

Finally, we consider the source neutrino flux from low-luminosity BL Lacs predicted in~\citet{Rodrigues:2020pli}. We adopt the same, negative source evolution shown in Fig.~1 of~\citet{Rodrigues:2020pli}, which results in the skymap shown in Fig.~\ref{fig:skymap_Rodrigues}.

\par
Since this model has the most negative source evolution, its resulting anisotropy level is the largest of the examples we consider here: $\alpha \simeq 0.759$. This is again in good agreement with our results from the main text. 

\par
In all three cases above the inferred anisotropy level, if measured, would result in a qualitatively accurate assessment of the underlying source evolution (e.g.\ positive vs negative). To obtain a quantitatively accurate measurement of the underlying source evolution a spectrum-specific anisotropy template should be used.

\bibliography{main}{}
\bibliographystyle{aasjournal}

\end{document}